\documentclass[preprint,10pt]{elsarticle}

\usepackage{amssymb}
\usepackage{bm}
\usepackage{inputenc}

\journal{Computer Science Review}

\begin{document}

\begin{frontmatter}



\title{Online Scheduling with Makespan Minimization: State of the Art Results, Research Challenges and Open Problems}


\author{Debasis Dwibedy\corref{cor1}\fnref{label1}}
\ead{debasis.dwibedy@gmail.com}
\author[label1]{Rakesh Mohanty}
\ead{rakesh.iitmphd@gmail.com}
\address[label1]{Department of Computer Science and Engineering, Veer Surendra Sai University of Technology, Burla 768018, Odisha,  India}

\begin{abstract}
Online scheduling has been a well studied and challenging research problem over the last five decades since the pioneering work of Graham with immense practical significance in various applications such as interactive parallel processing, routing in communication networks, distributed data management, client-server communications, traffic management in transportation, industrial manufacturing and production. In this problem, a sequence of jobs is received one by one in order by the scheduler for scheduling over a number of machines. On arrival of a job, the scheduler assigns the job irrevocably to a machine before the availability of the next job with an objective to minimize the completion time of the scheduled jobs.\\ 
This paper highlights the state of the art contributions for online scheduling of a sequence of independent jobs on identical and uniform related machines with a special focus on preemptive and non-preemptive processing formats by considering makespan minimization as the optimality criterion. We present the fundamental aspects of online scheduling from a beginner's perspective along with a background of general scheduling framework. Important competitive analysis results obtained by well-known deterministic and randomized online scheduling algorithms in the literature are presented along with research challenges and open problems. Two of the emerging recent trends such as resource augmentation and semi-online scheduling are discussed as a motivation for future research work. 

\end{abstract}


\end{frontmatter}


\section{Introduction}
Scheduling is a quintessential phenomenon in our daily life. Everyday we schedule meetings, set deadlines for projects, organize work periods, schedule games, manage time table for the lectures of various courses, allot rooms and plan maintenance operations. Each of these activities consists of a finite number of jobs. Our main objective is to finish the jobs in minimum possible time by designing an efficient schedule. When we have complete knowledge of all jobs at the outset and we make a schedule based on a sequence of available jobs on multiple machines, it is referred to as \textit{Offline Scheduling} [1-4]. However, in practical applications, all jobs are not known at the beginning. Mostly, jobs are available on the fly and upon receiving a job, the scheduler is constrained to assign the job irrevocably to one of the machines with no information on the future incoming jobs. Such a scheduling is known as \textit{Online Scheduling} [5]. Online scheduling has been well studied in the literature based on job characteristics, machine models, availability pattern of the jobs and optimality criteria [6].\\
\textbf{Motivation}. Offline $m$-machine($m\geq 2$) scheduling for \textit{makespan} minimization objective was shown to be $NP-Complete$ by a polynomial time reduction from the well-known Partition problem in [7-10]. Further, in online scheduling, the unavailability of all jobs at the beginning poses a non-trivial challenge in designing and analyzing the scheduling algorithms. However, \textit{online algorithm} [11, 12] provides a framework for processing a sequence of inputs which are given in an online fashion and the performance of such an algorithm is measured widely by \textit{competitive analysis} method [13]. For a basic understanding, we present briefly about online algorithm and competitive analysis as follows. 
\subsection{Online Algorithm and Competitive Analysis}\label{subsec: Online Algorithm and Competitive Analysis}
\textbf{Online algorithms} receive a sequence of inputs one by one and process each input irrevocably upon its availability to produce a partial output prior to the arrival of the next input. The partial output is produced by considering current and past inputs. Let us consider a sequence of inputs $ J=\left<J_1, J_2,.........J_n\right>$ of finite size $n$. At any time step $t$, the input sequence $<J_1, J_2...J_{t-1}, J_t>$ is received by an online algorithm, where $1 \leq t\leq n$. Each input $J_t$ is processed as soon as it is received with no information on future input sequence $<J_{t+1}, J_{t+2},..., J_{n-1}, J_n>$. Therefore, partial outputs $O_1, O_2,..., O_{n-1}$ are produced in an incremental way on the fly before the final output $O_n$ is obtained. In contrast, an offline algorithm receives the whole input sequence at the beginning and processes them simultaneously to produce the output $O$.\\
\textbf{Competitive analysis} method [8] measures the worst-case performance of any online algorithm $ALG$ by its \textit{competitive ratio}(CR), defined as the smallest positive integer $k(\geq 1)$, such that for all valid sequences of the inputs in set $J$= $\{J_1, J_2,...,J_n\}$, the following inequality holds \\
\hspace*{3.8cm}$C_{ALG}\leq k\cdot C_{OPT}$,\\
where $C_{ALG}$ is the \textit{makespan} obtained by online algorithm $ALG$ for any sequence of $J$ and $C_{OPT}$ is the optimum makespan incurred by the optimal offline algorithm $OPT$ for $J$. The \textit{Upper Bound}(UB) on the \textit{CR} obtained by $ALG$ guarantees the maximum value of \textit{CR}, for which the inequality holds for all legal sequences of $J$. The \textit{Lower Bound}(LB) on the \textit{CR} of any online problem $X$ ensures that there exists an instance of $J$ such that $ALG$ incurs makespan $C_{ALG}\geq b \cdot C_{OPT}$, where $b$ is referred to as \textit{LB} for $X$. The performance of $ALG$ is considered to be \textit{optimal} or \textit{tight} for any online problem $X$, when the \textit{UB} on the \textit{CR} achieved by $ALG$ for $X$ is also shown to be the \textit{LB} on the \textit{CR} of $X$. Sometimes, the performance of $ALG$ is referred to as \textit{tight} if $C_{ALG}$=$k\cdot C_{OPT}$ for all valid input sequences of $X$. The objective of an online algorithm is to obtain a \textit{CR} closer to the  \textit{LB} of the online problem or to obtain a \textit{CR} nearer to $1$. 
\subsection{Online Scheduling Problem }\label{subsec: Online Scheduling } 
We formally state the online scheduling problem with inputs, output, objective and constraints as follows.\begin{itemize} 
\item \textit{Inputs.} A list $M$=$(M_1, M_2,....M_m)$ of $m(\geq 2)$ machines, a sequence $J$=$<J_1, J_2,.....J_{n-1}, J_n>$ of $n(>>m)$ jobs are given, where each $J_i$ is characterized by its processing time $p_i$, where $1\leq i\leq n$.
\item \textit{Output.} Generation of a schedule, where \textit{completion time of the job that finishes last in the schedule} i.e. \textit{makespan($C_{max}$)} is the output parameter. 
\item \textit{Objective.} Minimization of the $C_{max}$. 
\item \textit{Constraints.} Jobs along with their processing times are revealed to the scheduler one by one in order; a newly available job must be scheduled irrevocably before the arrival of the next job; a job can not be forcefully preempted while it is in execution, but the preemptive variant of online scheduling supports the preemption of an ongoing job prior to its completion.
\end{itemize}
\subsection{Practical Applications}\label{subsec:Practical Applications} 
Online Scheduling has been intrinsically and widely used  in modern scientific and technical computations. In many real life scenarios, online scheduling has been found either as an independent problem or a segment of a larger problem. Major application  domains of online scheduling are mentioned in Table \ref{tab:Major Applications Areas of Online Scheduling}.
\begin{table}[!ht]
\caption{Major Applications of Online Scheduling}
\begin{tabular} {cp{6.6cm}}
\hline
\textbf{Domain(S)} & \textbf{Overview of Applications}  \\
\hline
Computers[6, 12, 14] & Interactive Parallel Processing, High Performance Computing, Computer based Simulation and Modeling, Robotics, Distributed Data Management\\
Networks[15] &  Routing, Client Request Management\\
Production and Manufacturing[16-17] & Production control System, Manufacturing Process Management \\
Transportation[18-19] & Traffic control and signaling\\
\hline 
\end{tabular}
\label{tab:Major Applications Areas of Online Scheduling}
\end{table}\\
We now briefly discuss about some important applications as follows. 
\subsubsection{Interactive Multi-Processing}\label{subsubsec:Interactive Multi-Processing}
An interactive multi-processing system receives jobs on the fly. Upon receiving a job, it has to immediately respond by allocating resource(s) such as memory or processing unit(s) to the job  with no knowledge of the future jobs. In practice, an interactive multi-processing system may be the operating system running on a parallel processing enabled computer, router of the communication networks, web server, robot navigation and motion planning system.
\subsubsection{Production and Manufacturing}\label{subsubsec:Production and Manufacturing}
Orders from clients arrive on the fly to a production system. The resources such as human beings, machinery equipment(s) and manufacturing unit(s) have to be allocated  immediately upon receiving each client order with no knowledge on the future orders. Online arrival of the orders have high impact on the renting and purchasing of the high cost machines in the manufacturing units.
\subsubsection{Traffic Management}\label{subsubsec:Traffic Management}
It is not known in advance the number of vehicles running on the road and passing through the traffic squares at any instance of time. For an effective transportation system, online scheduling can be very useful for managing traffic signals in various squares of a street or city.
\subsection{Scope of Our Survey}\label{subsec:Scope of Our Survey} 
This survey provides a comprehensive overview on the state of the art contributions for \textit{Online Scheduling} problem. In particular, the survey focuses on online $m$-machine scheduling with respect to \textit{machine models} such as identical and uniform related; \textit{job characteristics} such as preemptive and non-preemptive; \textit{optimality criterion} such as makespan(please, see section 2 to understand basic terminologies and notations related to the scope of our study).
The survey presents critical ideas, novel techniques along with important results obtained by seminal randomized and deterministic online scheduling algorithms to develop basic understanding from a beginner's perspective without discussing much detail on the proof techniques. To make the survey less exhaustive, some of the well studied areas such as online scheduling in real time systems, flow shop environment, unrelated machine, scheduling under machine availability or eligibility constraint, flow time objective and energy efficient scheduling are not covered.
\subsection{Uniqueness of Our Survey}\label{subsec:Uniqueness of Our Survey}
According to our knowledge, the existing literature lacks an exhaustive overview of the state of the art contributions of deterministic and randomized online scheduling algorithms. We present a summary of well-known surveys [6, 20-31] on online scheduling in Table \ref{tab:Relatedsurvey}. 
\begin{table}
\centering
\caption{Well-known Related Surveys on Online Scheduling}
\label{tab:Relatedsurvey}
\begin{tabular} {ccp{7.3cm}}
\hline
\textbf{Year} & \textbf{Author(s)} & \textbf{Scope/Main Contributions} \\
\hline
1977 & Gonzalez [20] & Study of scheduling in uni-processor, multiprocessor and job shop environment.\\
1979 & Graham et al. [6] & Classification of scheduling problem in  standard framework of $\alpha|\beta|\gamma$. Overview of deterministic strategies for scheduling in various machine models such as single, uniform related and unrelated machines, flow and open shop scheduling. \\
1989 & Cheng et al. [21] &  Job shop scheduling with deadline assignment. \\
1998 & Sgall [22] & Scheduling strategies for variants of online scheduling by considering release time, preemption and precedence constraints, speeds of machines and shop scheduling.  \\
1998 & Chen et al. [23] & Methodologies, complexity analysis in non-preemptive and preemptive online scheduling for unit length jobs. Study of shop scheduling, family scheduling, resource constraint scheduling and scheduling with communication delays. \\  
2004 & Pruhs et al. [24] & Flow time minimization in single and parallel machines. \\
2006 & Brucker [25] & Shop scheduling, scheduling with deadline, batch scheduling and multi-purpose machine scheduling. \\
2008 & Leung et al. [26] &  Offline and online scheduling with processing set restrictions in non-preemptive, preemptive settings for makespan minimization.\\
2009 & Albers [27] & Energy efficient online scheduling. Specific results in load balancing, makespan and flow time minimization. \\
2012 & Sgall [28] & Open problems in online scheduling for the maximization of the throughput. \\
2013 & Lee et al. [29] & Online $m$-machine scheduling under machine eligibility constraint.\\
2018 & Epstein [30] & Semi-online scheduling for makespan minimization in identical and uniform related machines.\\
2019 & Beaumont et al. [31] & Scheduling in a parallel computing system with heterogeneous resources. Design of a framework to compare the performances of related algorithms based on achieved makespans and time complexities. \\ 
\hline
\end{tabular}
\label{tab:Relatedsurvey}
\end{table}    
This survey studies the fundamental aspects of online  computation by a historical chronological overview on the seminal contributions for preemptive and non-preemptive online scheduling in identical and uniform related machines. Fourteen well-known online scheduling algorithms are presented. Deterministic and randomized online scheduling algorithms for makespan minimization are chronologically described for developing a basic understanding. Major research issues, open problems and two of the emerging recent trends are briefly presented as a motivation for future research work.
\section{Preliminaries}\label{subsec:Preliminaries}
We present basic terminologies, notations and definitions which are used in our survey in Table \ref{tab:Basic Terminologies Notations and Definition}. Based on the scope of our literature survey, we make a classification of the online scheduling problem by three parameters such as parallel machine models, job characteristics and optimality criterion. We now briefly discuss about the parameters as follows. 
\begin{table}[!htbp]
\caption{Basic Terminologies Notations and Definitions}
\begin{tabular} {ccp{6.6cm}}
\hline
\textbf{Terms} & \textbf{Notations} &\textbf{Definitions/Descriptions} \\
\hline
Job[1] & $J_i$ &  Program under execution which  consists of a finite number of instructions. A job is also referred to as a collection of at least one smallest indivisible sub task called \textit{thread}. We use terms job and task in the same sense. \\
Processing Time[1-3] &  $p_{ij}$ & Total time of execution of a job $J_i$ on a machine $M_j$. For identical machines processing time of $J_i$ is denoted by $p_i$.\\
Release Time[2, 25] & $r_i$ & Time at which any job $J_i$ becomes available for processing.  \\
Completion Time[3, 25] & $c_i$ & The time at which any job $J_i$  finishes its execution\\
Deadline[3, 25] & $d_i$ & Time by which job $J_i$ must be completed.\\
Machine[1] & $M_j$ & An automated system capable of processing some tasks by following a set of rules.  We use terms machine and processor in the same sense. \\
Load [11] & $l_j$ & Sum of processing times of the jobs that have been assigned to machine $M_j$. \\
Speed [2, 3] & $S_j$ & The number of instructions processed by the machine in unit time \\
Speed Ratio & $s$ & The ratio between the speeds of two machines. For $2$-machines with speeds  $1$ and $\frac{1}{S}$ respectively. We have speed ratio: $s$ = $\frac{1}{\frac{1}{S}} = S$\\
\hline
\end{tabular}
\label{tab:Basic Terminologies Notations and Definition}
\end{table} 
\subsection{Parallel Machine Models}\label{subsec: Parallel Machine Models}
In parallel machine system, all threads of a job execute simultaneously in a set of machines $\mu$ such that $\mu\subseteq\{M_1,.........M_m\}$. Here, we consider parallel processing of multiple jobs, by assuming that each job consists of a single thread of execution. We consider parallel machine models such as identical and uniform related machines in our survey, which are presented as follows. 
\begin{itemize}
\item \textit{Identical Machines}: Here, all machines have equal speeds of processing a job $J_i$ such that $p_{ij}=p_i$, $\forall M_j, 1\leq j \leq m$ and $p_i\geq 1$. 
\item \textit{Uniform Related Machines or Uniform Machines}: Here, the speeds of the machines may differ from one another. For a uniform machine $M_j$ with speed $S_j$, the execution time of job $J_i$ is $p_{ij}=\frac{p_i}{S_j}$ .
\end{itemize}
\subsection{Job Characteristics} \label{subsec: Job Characteristics}
Job characteristics describe about the nature of the jobs and various constraints related to job scheduling.
We consider the following job characteristics.
\begin{itemize}
\item \textit{Preemption.} It allows splitting of a job into pieces, where each piece is executed on same or different machines in non-overlapping time intervals.
\item \textit{Non-preemption.} It ensures that once a job $J_i$ with processing time $p_i$ begins to execute on machine $M_j$ at time $t$, then $J_i$ continues the execution on $M_j$ till time $t+p_i$ with no interruption. 
\item \textit{Precedence Relation.} It defines dependencies among the jobs by the partial order '$\prec$' rule on the set of jobs [5]. A partial order can be defined on two jobs $J_i$ and $J_k$ as $J_i$ $\prec$ $J_k$, which means execution of $J_k$ never starts before the completion of $J_i$. The dependencies among different jobs can be illustrated with a precedence graph $G(p, \prec)$, where each vertex represents a job and labeled with its processing time $p$. A directed arc between two vertices in $G(p, <)$ i.e $J_i$ $\rightarrow$ $J_k$ represents $J_i$ $\prec$ $J_k$, where $J_i$ is referred to as \textit{predecessor} of $J_k$. If there exists a cycle in the precedence graph, then scheduling is not possible for the jobs. When there is no precedence relation defined on the jobs, then they are said to be \textit{independent}.
\end{itemize}
\subsection{Optimality Criterion}
\label{subsec: Optimality Criterion}
We consider makespan $(C_{max})$ as the optimal criterion. Makespan is defined as completion time of the job that finishes last in a schedule. Formally, $C_{max}=\max_{1 \leq i \leq n}{c_i}$. The objective is to obtain minimum $C_{max}$. Another way of interpreting makespan is in terms of \textit{load balancing($L_{max}$)}. Here, the makespan is defined as the largest load incurred on any machine $M_j$. Formally, $L_{max}=\max_{1 \leq j \leq m}{l_j}$. Here, the objective is to obtain minimum $L_{max}$.
\section{Well-Known Online Scheduling Algorithms}
\label{sec: Well-Known Online Scheduling Algorithms} 
We present ten deterministic and four randomized online scheduling algorithms for identical and uniform related machines as follows.
\subsection{Deterministic Algorithms} \label{subsec: Deterministic Algorithms}
Deterministic algorithms[11-12, 32] obtain same output while processing a given input a number of times by following each time the same sequence of steps. Here, the output and running time depend on input(s) only. We now present ten well-known deterministic online scheduling algorithms as follows.
\begin{itemize}
\item \textbf{Algorithm List Scheduling(LS)} was proposed by Graham [5] for non-preemptive online scheduling on identical machines. Algorithm \textit{LS} assigns an incoming job $J_i$ to the machine $M_j$, which is least loaded among the machines. 
\item \textbf{Algorithm Largest Processing Time(LPT)} was proposed by Graham [33] for non-preemptive off-line scheduling on identical machines. \textit{LPT} first orders a list of jobs in non-increasing sizes. Then, it assigns job one by one from  the ordered list to the machine, which has least load after each assignment of a job. \textit{LPT} is an offline algorithm, however it has been used significantly as an intermediate step in many online scheduling algorithms. 
\item \textbf{Algorithm Refined List Scheduling(RLS)} was proposed by Galambos and Woeginger [34] for non-preemptive online scheduling on $m$-identical machines. Upon the arrival of any job(say $J_i$), algorithm \textit{RLS} first orders the machines in non-decreasing loads. $J_i$ is assigned to $M_1$ if $l_1 \leq \alpha l_m$ and $J_i$ is assigned to $M_2$ if  $l_1+p_i > \alpha l_m$, where $0 \leq \alpha \leq 0.33$. There after, \textit{RLS} assigns next incoming jobs ($J_{i+1}$) to $M_1$ till the following inequality holds: $\frac{l_1}{\beta} \leq p_{i+1} \leq \beta(l_m)$, where $1 \leq \beta \leq 1.25$. Then, the machines are re-ordered again.
\item \textbf{Algorithm ASSIGN-2:} Aspnes et al. [15] proposed the algorithm \textit{ASSIGN-2} for non-preemptive online scheduling of uniform related machines. \textit{ASSIGN-2} works in phases by guessing the cost of \textit{OPT} at each phase. Upon receiving the first job $J_1$, the cost of \textit{OPT} is initialized to the load incurred by $J_1$ on the fastest machine. The cost of \textit{OPT} remains same till the completion of a phase. Each incoming job $J_i$ is assigned to the machine $M_j$, which has slowest speed among all machines and $l_j+p_i \leq 2(OPT)$. Subsequently, the load of $M_j$ is updated  and the machines are ordered in non-decreasing speeds.  The algorithm ends a phase, when it does not find an appropriate $M_j$. A new phase is started by initializing the loads of all machines to $0$ and by doubling the previous value of \textit{OPT}. Then, $J_i$ is assigned to the slowest machine $M_j$ for which $l_j+p_i \leq 2(OPT)$.
\item \textbf{Algorithm Chen Vliet Woeginger(CVW)} was proposed by Chen et. al. [35] for preemptive online scheduling of $m$-identical machines. Originally, there was no particular name given to the algorithm. We name the algorithm as \textit{CVW} by extracting the first letter of the author's names. For each incoming job $J_i$, algorithm \textit{CVW} computes the cost of \textit{OPT}, which is $\max\{\frac{1}{m}\sum_{i=1}^{n}{p_i},\hspace{0.1cm} \max_{1 \leq i \leq n}{p_i}\}$. Then, job $J_i$ is assigned to any $M_j$, where $1 \leq j \leq m$, if the new load of machine $M_j$ is at most ($\frac{m^m}{m^m-(m-1)^m}$) times of \textit{OPT}. Otherwise, sequence the machines in non-decreasing loads and assign ($\frac{m^m}{m^m-(m-1)^m})OPT$-$l_1$ portion of job $J_i$ to machine $M_1$ and the rest part of job $J_i$ to machine $M_2$. 
\item \textbf{m-Machine Algorithm} was proposed by Bartal et al. [36] for non-preemptive online scheduling of $m$-identical machines. The algorithm first orders the machines in non-decreasing loads such that $M_1$ has least load and $M_m$ has highest load. When the first job $J_i$ is available, it is assigned to $M_1$ and the machines are ordered in non-decreasing loads. Then, logically group $m$-machines in to two, where first $\sigma m$ machines constitute first group and rest $m-\sigma m$ machines form the second group. Each incoming job $J_{i+1}$ is assigned to ${M_{\sigma m +1}}^{th}$ machine, if $l_{\sigma m +1}$ after the assignment of $J_{i+1}$ is at most $1.985$ times of the average load incurred on first $\sigma m$ machines. Otherwise, $J_{i+1}$ is assigned to $M_j$, which has minimum load over all machines. Note that, after each assignment of a job, the machines are ordered in non-decreasing sequence of their loads. In the former case, machines from $M_{\sigma m+1}$ to $M_{m}$ are to be ordered and in the later case all machines from $M_1$ to $M_m$ are to be ordered. 
\item\textbf{ Algorithm Compare Height to Average of Shorter Machines (CHASM)} was proposed by Karger et al. [37] for non-preemptive online scheduling in identical machines. \textit{CHASM} always aims to maintain a light load on the first $k$ machines by keeping the next $m-k$ machines with heavy load. Algorithm \textit{CHASM} works in the following way: upon the arrival of a new job $J_i$, it first orders $m$-machines in non-decreasing sequence of their loads and  assigns $J_i$ to the $({k+1})^{st}$ least loaded machine $M_{k+1}$ if $l_{k+1}+p_i \leq 1.945(\frac{1}{k}\sum_{j=1}^{k}{l_j})$, where $1 \leq k < m$.  Otherwise, assign $J_i$ to the most lightly loaded machine $M_1$. 
\item \textbf{Algorithm H} was proposed by Wen and Du [38] for preemptive online scheduling of $2$-uniform related machines. Algorithm \textbf{H} works in the following way: let the current loads of machines $M_1$ and $M_2$ be $l_1$ and $l_2$ respectively. Upon receiving a new job $J_i$, Algorithm \textbf{H} first splits $J_i$ in to two parts. The largest part of $J_i$ which is of size at most $\alpha.C^{i-1}_{OPT}-l_2$ is assigned to the fastest machine $M_2$, where $S_2 \geq 1$. The remaining part of $J_i$, which is of size at most $l_2-l_1$ is assigned to the machine $M_1$, where $S_1=1$. Here, $C^{i-1}_{OPT}$ is the cost of \textit{OPT} before the assignment of $J_i$. We have $\alpha=\frac{({1+S}^2)}{1+S+S^2}$.
\item \textbf{Algorithm $M_2$} was proposed by Albers [39] for non-preemptive online scheduling on $m$-identical machines. Algorithm $M_2$ always maintains $m$-machines, which are numbered in non-decreasing order of their loads after the assignment of each incoming job $J_i$ such that $M_1$ be the machine with minimum load and $M_m$ be the machine with maximum load. A new job $J_i$ is assigned to machine $M_1$ if one of the following inequalities holds.\\
(a)\hspace* {0.1cm} $L_l \leq \alpha(L_h)$ \hspace* {0.3cm} (b) \hspace* {0.1cm} $\lambda_{m} > l^{i-1}_m$ \hspace* {0.1cm} and \hspace* {0.1cm} $\lambda_{m} > a.\frac{L_l+L_h}{m}$. Otherwise, $J_i$ is assigned to machine $M_{k+1}$. Here, we have the values of $a=1.923$, $b=0.29m$, $k=\lfloor\frac{m}{2}\rfloor$ \hspace* {0.1cm} and \hspace* {0.1cm} $\alpha=\frac{(a-1)k-\frac{b}{2}}{(a-1)(m-k)}$.  If $J_i$ is scheduled on $M_1$, then we have $L_l=\sum_{i=1}^{k}{l_i}$ \hspace* {0.1cm} and \hspace* {0.1cm} $L_h=\sum_{i=k+1}^{m}{l_i}$. If $J_i$ is scheduled on $M_{k+1}$, then we have $\lambda_{m}=\max_{1 \leq j \leq m}{l_j}$ \hspace* {0.1cm} and \hspace* {0.1cm} $l^{i-1}_{m}$ is the load of most loaded machine $M_m$ before the assignment of $J_i$.
\item \textbf{Algorithm MR} was proposed by Fleischer and Wahl [40] for non-preemptive online scheduling of $m$-identical machines. Algorithm MR schedules a sequence of jobs in the following way: initially, sort all machines in non-decreasing order of their loads such that $M_1$ and $M_m$ are the least loaded and most loaded machines respectively. When ever a new job $J_i$ arrives, assign $J_i$ to the machine $M_{k+1}$ if $\lambda > \alpha(l_{2k+1})$ and $l_{k+1}+p_i \leq a(L^i_{avg})$. Otherwise, schedule $J_i$ on machine $M_1$. After each assignment of a new job, the load of the corresponding machine is updated and all machines are re-sorted. Here, we have $k \approx \lfloor0.36m \rfloor+1$, $\alpha \approx 0.46$, $\lambda=\frac{1}{k}\sum_{i=1}^{k}{(l_i)}$, which is the average load of $k$ least loaded machines before the assignment of $J_i$ and  $L^i_{avg}$ be the average load incurred on $m$-machines after assigning $J_i$. 
\end{itemize}
\subsection{Randomized Algorithms} \label{subsec: Randomized Algorithms}
A randomized algorithm [41] flips coin while processing a given input. The algorithm produces a different output or follows a different order of execution steps at each run. Here, the output and running time depend on the input(s) and random bits. We now present some well-known randomized online scheduling algorithms as follows.  
\begin{itemize}
\item \textbf{Algorithm Rand-2} was proposed by Bartal et al. [36] for non-preemptive online scheduling on $2$-identical machines. Algorithm \textit{Rand-2} maintains two logical schedules for each incoming job $J_i$. In the first schedule, $J_i$ is assigned to the least loaded machine and in the second schedule $J_i$ is allocated on the most loaded machine. Overall expected \textit{discrepancies} $E_1$ and $E_2$ are computed for both the schedules respectively, where \textit{discrepancy} is the difference in loads of the two machines at any instance of time. Then  a value for $x$ is chosen, where $0 \leq x \leq 1$ such that $x(E_1)+(1-x)E_2 \leq \frac{1}{3}L$, where $L$ is the total load incurred by jobs $J_1, J_2, ...J_{i-1}\hspace* {0.1cm} and \hspace* {0.1cm} J_i$. If such a value of $x$ exists, then $J_i$ is actually scheduled on the least loaded machine with probability $x$ and $J_i$ is assigned to the most loaded machine with probability $1-x$. If there exists no such a value of $x$, then schedule job $J_i$ explicitly on the least loaded machine. 
\item \textbf{Algorithm Linear Invariant(LI)} was proposed by Seiden [42] for non-preemptive online scheduling on $m$-identical machines, where $2 < m \leq 7$. Algorithm \textit{LI} schedules each incoming job $J_t$ on machine $M_1$ with probability $p$ and on machine $M_2$ with probability $1-p$, where $p=\frac{l^{2}_{m}-\alpha(l^{2}_{1})}{l^{2}{m}-\alpha(l^{2}_{1})-(l^{1}_{m}-\alpha(l^{1}_1))}$. Here, $l^{2}_{1}$ and $l^{2}_{m}$ are the loads of current least loaded machine $M^{2}_1$ and most loaded machine $M^{2}_{m}$ respectively, when all jobs $J_i$, where $1 \leq i \leq t$ are scheduled each time on the \textit{second least loaded machine} $M^{2}_{2}$. We have $l^{1}_{1}$ and $l^{1}_{m}$ as the loads of current $M^{1}_{1}$ and $M^{1}_{m}$ respectively, when all $J_i$ are assigned each time to the least loaded machine $M_1$. We have values of $\alpha$ equals to 1.80, 2.04, 2.12, 2.11 and 2.10 for $m$=3,4,5,6,7 respectively. It is observed that after each assignment of a job, machines are re-indexed in non-decreasing order of their loads.
\item \textbf{Algorithm BIAS} was proposed by Epstein et al. [43] for non-preemptive online scheduling on $2$-uniform-related machines, where speed of machine $M_1$ is $S_1=1$ and speed of machine $M_2$ is $S_2 \geq 1$. Algorithm \textit{BIAS} schedules an incoming job $J_i$ on the fastest machine $M_2$ with probability $\frac{S}{2}$, otherwise assigns $J_i$ to the slowest machine $M_1$.
\item \textbf{Algorithm RAND} was proposed by Albers [44] for non-preemptive online scheduling on $m$-identical machines. Algorithm \textit{RAND} is basically a combination of two deterministic algorithms ${ALG}_1$ and ${ALG}_2$. Any input job stream $\sigma$ is scheduled by algorithm ${ALG}_1$ with probability $P=\frac{1}{2}$ and by algorithm ${ALG}_2$ with probability $1-P$.\\
\textit{Algorithm ${ALG}_1$:} Machines are always indexed in the non-decreasing order of their loads. A new job $J_i$ is scheduled on machine $M_{k_1+1}$ if the schedule is \textit{critical} and $l^{i-1}_{1, k_1+1}+p_i \leq b_1(\frac{L^{i}}{m})$. Otherwise, $J_i$ is assigned to the least loaded machine $M_1$. Here, $l^i_{x,j}$ be the load of the machine $M^i_{x,j}$, which is the $j^{th}$ current least loaded machine after scheduling jobs from $J_1$ to $J_i$ by algorithm ${ALG}_x$, where $x \in \{1, 2\}$. $L^i$ be the total processing time incurred by the jobs from $J_1$ to $J_i$. A schedule becomes \textit{critical} if ${\mu}^{i}_{x} > {\alpha}_x(L^i_x), 2k_x+1$, where ${\mu}^i_x=\frac{1}{k_x}\sum_{j=1}^{k_x}{l^i_{x, j}}$. The values of other parameters are initially set as follows: $b_1=1.832$, $k_1=\lceil\frac{9}{25}m\rceil$, $\alpha_1=1-\frac{k_1-\lfloor 0.074m\rfloor}{2-0.916k_1}$.\\
\textit{Algorithm ${ALG}_2$:} Upon receiving a new job $J_i$, algorithm ${ALG}_2$ first runs algorithm ${ALG}_1$. If the schedule obtained after assigning $J_i$ by algorithm ${ALG}_1$ is balanced, meaning the machines are now equally loaded, then algorithm ${ALG}_2$ sets the value of $\gamma$ to $\max\{b'_2\frac{L^i}{m}, \hspace* {0.1cm} b_2\beta^{-1}\frac{\lambda^i_1}{m}\}$, where  $\lambda^i_1=\sum_{j=k_1+1}^{m}{l^i_{1, j}}$. Otherwise,  the value of $\gamma$ is set to $b'_2\frac{L^i}{m} $. Now, algorithm ${ALG}_2$ assigns $J_i$ to the machine $M_{k_2+1}$ if the schedule is critical and $l^{i-1}_{1, k_2+1}+p_i \leq \gamma$. Otherwise, $J_i$ is scheduled on the least loaded machine $M_1$. Initialize the values of other parameters as follows: $b_2=2$, $b'_2=1.885$, $k_2= \lceil0.375m\rceil$, $\alpha_2=0.449$ and $\beta=1-(b_1-1)\frac{k_1}{m}$. Note that: after each assignment of a new job, the machines are re-numbered according to the non-decreasing order of their loads. 
\end{itemize}
\section{Historical Overview of Online Scheduling}
\label{sec: Historical Overview of Online Scheduling}
The theory of sequencing and scheduling have been emerging as an interesting area of research over the past few decades. In early fifties, the main focus of research was on offline single machine scheduling. After a decade, the curiosity was transferred to define potential advantages of multiprocessing systems. As an outstanding outcome, the systems witnessed increase in throughput. Still, the quest was for designing application specific scheduling models and to obtain optimum schedule for processing of multiple jobs. This resulted in the emergence of a number of scheduling setups. One of such setups is the \textit{online scheduling}, which basically deals with the \textit{online arrival} of a sequence of jobs. Online Scheduling setup was first proposed and validated by Graham in 1966 [5], however it has gained significant research interests after the introduction of \textit{competitive analysis} in 1985 [13]. We now present an overview of the early contributions on online scheduling from year 1966 to 1984 as follows.\\     
Graham [5] initiated the study of non-preemptive online scheduling on $m$-identical machines in the objective to explore several multiprocessing timing anomalies. He considered a sequence of $n(\geq 2)$ jobs and ordered them through a \textit{static list}, where the ordering of the jobs is decided upon receiving all jobs. Here, the jobs are ordered only once and the ordering remains same throughout the scheduling of all jobs. Graham proposed the famous \textit{LS} algorithm, which always scans the list until an \textit{eligible} job is found. A job $J_i$ is called \textit{eligible}, if execution of $J_i$ has not been started and processing of all its predecessors have been completed. Algorithm \textit{LS} adopts a \textit{greedy} strategy by immediately scheduling the first eligible job of the list to the lowest loaded machine prior to make another scan of the list. Algorithm \textit{LS} achieves a \textit{performance ratio} of $2-\frac{1}{m}$, which is the ratio between the largest makespan obtained over all possible ordering of the jobs to the \textit{optimum makespan($C_{OPT}$)}, where $C_{OPT}$=$\frac{1}{m}\cdot (\sum_{i=1}^{n}{p_i})$. In [33], Graham proposed \textit{LPT} algorithm, which orders a sequence of jobs in a \textit{dynamic list}, where the ordering of the jobs may change during the scheduling process. Algorithm \textit{LPT} always places the eligible job with largest size among the available jobs in the top of the list and schedules it by algorithm \textit{LS}. Algorithm \textit{LPT} achieves a performance ratio of $1.33-\frac{1}{3m}$. These two studies of Graham showed a new direction in scheduling, which is based on the \textit{arrival pattern} of the jobs. The scenario of selecting the eligible job from the top of the list without looking at the entire task list was later formed the concept of arrival of jobs \textit{one by one}.\\
Coffman and Graham [45] followed the work of Graham [5,33] and designed two non-enumerative algorithms with worst case complexities of $n^2$ and $\frac{1}{4}(n^2)$ respectively for makespan minimization of a job schedule. They considered a list of jobs, where all jobs have equal execution time. The main objective of both algorithms is to obtain a list $A^*$ such that makespan of $A^*$ is minimal over all  list $A$.  A novel computational model was introduced, which partitions a given problem into two equal sized jobs. The applicability of the model was shown in preemptive scheduling, where a job can be splitted and shared among multiple machines.\\
Sahni [46] studied scheduling of $n$-independent jobs on $m$-identical machines for minimizing the makespan. He designed an offline algorithm by \textit{dynamic programming} approach, which has the worst case complexity of $O(min\{2^n, nM\})$, where $M$ characterizes the cost of \textit{OPT}. Furthermore, he proposed three approximation algorithms for scheduling problems such as single machine scheduling with deadline, scheduling on $m$-identical machines for minimizing completion time and scheduling on $2$-identical machines to minimize weighted mean flow time. Here, all proposed algorithms obtain costs which are not far than a value $\epsilon$ from the optimum cost, where $0 < \epsilon < 1$. The worst case complexities of both algorithms were proved to be $O(\frac{n^3}{\epsilon})$.\\
Sahni and Cho [47] proposed a nearly online algorithm for preemptive scheduling of $n$-independent jobs on $m$-uniform related machines. They specified each job $J_i$ with release time $r_i$. The worst case time complexity of the algorithm was proved to be $O(m^2n+mnlogn)$. The algorithm ensures at most $O(nm)$ preemption in executing all jobs. Here, they assumed that there are at most $v$ distinct release times of the jobs and designed an algorithm which has $v$ phases. At any phase $k$, where $1 \leq k < v$, a selected number of jobs are scheduled through a deterministic procedure. In fact, those jobs are chosen which have non-zero remaining processing times and are available on or before time $r_k$. In the last phase, the jobs are scheduled in the interval $[r_v, d]$, where $d$ is assumed to be the common deadline for all jobs.\\
Hariri and Potts [48] proposed a \textit{branch and bound} algorithm to obtain a processing order of the jobs, which minimizes the sum of weighted completion times in a single machine offline scheduling environment. They considered a list of jobs, where each job is specified by its release time and weight. They computed earliest time of completion for each job. They obtained a lower bound by assuming \textit{Langragean} relaxation for the release time constraints. Here, \textit{Langragean multipliers} are chosen in such a way that the generated job sequence yields optimum cost for the relaxed problem. Later, they provided a method to derive better release time constraints than the original ones to increase the lower bound. Blazewicz et al.[49] studied parallel machine scheduling of a sequence of unit size jobs under various resource constraints schemes. \\
Initial two decades(1966-1985) were focused mostly on exploring variants of the Graham's \textit{List Scheduling} setup in optimizing the makespan. \textit{Greedy, Dynamic programming}, \textit{branch and bound} algorithmic design strategies were introduced for the multiprocessor scheduling problem. A more systematic study on online $m$-machine scheduling was started after the development of \textit{competitive analysis}. Hence forth, two of the major aspects of online algorithm design such as deterministic and randomized strategies were studied extensively. We present the state of the art on deterministic and randomized online scheduling in the coming sections.
\section{Deterministic Online Scheduling: State of The Art}\label{sec:Deterministic Online Scheduling: State of The Art}
This section presents important results with insights and research challenges in the design of deterministic online  algorithms for preemptive and non-preemptive online scheduling with makespan minimization for identical and uniform- related machines. 
\subsection{ Non-preemptive, Identical Machines}
\label{subsec: Non-preemptive, Identical Machines}
\textbf{Upper Bound Results.} Graham's \textit{LS} [5] algorithm was provably the first deterministic online algorithm for $m$-machine scheduling problem. The objective of algorithm \textit{LS} is to assign jobs to the machines such that at the end of scheduling, all machines incur nearly equal load. However, \textit{equal load sharing} policy fails to obtain optimum makespan in all cases. One such case is the availability of the largest job as the last  job i.e. $n^{th}$ job of the sequence while the assignment of first $n-1$ jobs have incurred nearly equal load on all machines. In this case, algorithm \textit{LS} obtains a makespan($C_{LS}$) which is nearly twice of the value of the optimum makespan($C_{OPT}$). Let us consider a list of $m^2-m+1$ jobs, where a sequence of $m(m-1)$ jobs each of \textit{size} $1$ unit arrive one by one and a job of \textit{size} $m$ unit is available at the end of the sequence. We now have $C_{LS}\geq 2m-1$, where  $C_{OPT}$=$\frac{(m^2-m)\cdot 1 + m}{m}$=$m$. Therefore, we have $\frac{C_{LS}}{C_{OPT}}\geq 2-\frac{1}{m}$. An equal \textit{UB} was shown by Graham to obtain ($2-\frac{1}{m}$)-competitiveness of algorithm \textit{LS}. It was a non-trivial challenge to design improved competitive online algorithm with \textit{CR} asymptotically lesser than $2$.  After a quest of over two and half decades, the first improvement over algorithm \textit{LS} was presented by Galambos and Woeginger [34]. They proposed algorithm \textit{RLS} and proved a \textit{UB} of $2-\frac{1}{m}$ - $\epsilon_m$, where $\epsilon_m > 0$ for $m \geq 4$. 
Bartal et al. [36] proposed the \textit{m-Machine Algorithm} and achieved \textit{UB} $1.985$ for $m \geq 70$. The algorithm assigns incoming jobs to the machines in such a way that there exists always a set of machines with light load and rest machines with heavy load. So, whenever a job arrives, it can be assigned to the lightly loaded machine of the set of heavily loaded machines or to the smallest loaded machine among all machines. The objective is to obtain a makespan which is smaller than twice the value of the optimum makespan.
Karger et al. [37] proposed the algorithm \textit{CHASM} and obtained an improved \textit{UB} $1.945$ for all $m$. Algorithm \textit{CHASM} out performs algorithm \textit{LS} for $m \geq 6$. An improvement over \textit{CHASM} was proposed by Albers [39]. She designed the \textit{Algorithm $M_2$} and proved a \textit{UB} of $1.923$ for a general case of $m$-machine. Fleischer and Wahl [38] proposed the algorithm \textit{MR}, which obtains a \textit{UB} of $1.9201$ for $m\rightarrow \infty$. Algorithm \textit{MR} is currently the best deterministic pure online algorithm for non-preemptive online  scheduling on $m$-identical machines for makespan minimization, where $m \geq 64$.\\
\textbf{Lower Bound Results.}
Faigle et. al. [50] proved that \textit{LS} is optimal online scheduling algorithm for $m=2,3$. The \textit{LB} $2-\frac{1}{m}$ for $2$-machine case was shown by considering online availability of a sequence of three jobs $<J_1, J_2, J_3>$ with $p_1$=$1$, $p_2$=$1$ and $p_3$=$2$. Similarly, for $m=3$, they showed \textit{LB} $2-\frac{1}{m}$ by considering online arrival of a sequence of seven jobs, where the jobs are of sizes ($1,1,1,3,3,3,6$) respectively. Further, they proved \textit{LB} $1.707$ for $m \geq 4$ by considering a sequence of $2m+1$ jobs, where first $m$ jobs are of size $1$ unit each, next $m$ jobs are of size $1+\sqrt{2}$ unit each and the last job is of size $2(1+\sqrt{2})$ unit. \\
Bartal et al. [51] obtained \textit{LB} $1.837$ for $m \geq 3454$ by analyzing a special class of input job sequence. They considered a sequence of $4m+1$ jobs, where first $m$ jobs are of size $\frac{1}{x+1}$ unit each, second $m$ jobs are of size $\frac{x}{x+1}$ unit each, third $m$ jobs are of size $x$ unit each, next $\lfloor\frac{m}{2}\rfloor$ jobs are of size $y$ unit each, next $\lfloor\frac{m}{3}\rfloor-2$ jobs each with size $z$ unit and last ($m+3-\lfloor\frac{m}{2}\rfloor-\lfloor\frac{m}{3}\rfloor$) jobs are of size $2y$ unit each, where $x,y,z$ are positive real values. Albers [39] obtained a better \textit{LB} of $1.852$ for $m\geq 80$. She considered a special class of job sequence, where jobs are available in four rounds. In each round a specific number of equal size jobs arrive. The main idea is to schedule the jobs round-wise such that the makespan incurred at each round is not more than $1.852$ times of the value of $C_{OPT}$. Gormley et al. [52] improved the \textit{LB} to $1.853$ for $m \geq 80$ by considering an adversary strategy. The adversary strategy was presented as a game tree. Here, the game tree has two kinds of nodes, the adversary request nodes and the online move nodes. Each adversary request node is a non-leaf node. The children of adversary request nodes are online move nodes, which are generated at each move as per the current request. Each online move node is either a leaf or has a single adversary request node as a child. \\
Rudin III [53] proved that no deterministic online algorithm can achieve a competitive ratio smaller than $1.88$. Later, he along with Chandrasekaran [54] obtained \textit{LB} $\sqrt{3}$ for $m=4$. For the general case of $m$-machine, they showed \textit{LB} $\sqrt{3} - \epsilon$, where $\epsilon$ is a positive constant. They used the job master strategy to produce successive layers of jobs, where each layer contains $m$ jobs. The layers are considered in such a way that if any two of the jobs in the same layer are scheduled on the same machine, then the makespan of the corresponding machine will become at least $\sqrt{3} - \epsilon $ times of the value of the $C_{OPT}$. Therefore, irrespective of the scheduling algorithm, by the completion of a series of layers, it can be known that either there has been assignment of one job to each machine at each layer or a competitive ratio of at least $\sqrt{3} - \epsilon$ has already been achieved. We now present the summary of the important competitive analysis results achieved by deterministic online algorithms for non-preemptive scheduling on identical machines in Table \ref{tab:Summary of Important Results for Identical Machines by Deterministic Algorithms}. 
\begin{table}[!ht]
\centering
\caption{Summary of Important Competitive Analysis Results}
\begin{tabular} {ccp{2.9cm}c}
\hline
\textbf{Year} & \textbf{Author(s)} & \textbf{Result(s)} & \textbf{Bound} \\
\hline 
1966 & Graham[5] &  $2-(\frac{1}{m})$, $m$=$2, 3$ & UB.\\
1993 &  Galambos, Woeginger[34] &  $2-(\frac{1}{m} - \epsilon_m)$, $m \geq 4$ & UB.\\
1995 &  Bartal et al.[36] &  $1.986$, $m \geq 2$ & UB.\\
1996 &  Karger et al.[37] &  $1.945$, $m \geq 8$ & UB.\\
1999 &  Albers[39] &  $1.923$, $m \geq 2$ & UB.\\
2000 &  Fleischer, Wahl[40] &  $1.9201$, $m\rightarrow \infty$ & UB.\\
1989 & Faigle et al. [50] &  $1.707$, $m \geq 4$ & LB.\\
1994 &  Bartal et al. [51] &  $1.837$, $m \geq 3454$ & LB.\\
1999 &   Albers[39] &  $1.852$, $m \geq 80$ & LB.\\
2000 &  Gormley et al. [52] &  $1.85358$, $m \geq 80$ & LB.\\
2001 &  Rudin III [53] &  $1.88$, $m \rightarrow \infty$ & LB.\\
2003 &  Rudin III, Chandrasekaran [54] &  $\sqrt{3} - \epsilon$, $m=4$ & LB.\\
\hline
\end{tabular}
\label{tab:Summary of Important Results for Identical Machines by Deterministic Algorithms}
\end{table}\\
\textbf{Research Challenges:} 
\begin{itemize}
\item Minimizing or diminishing the gap between current best \textit{LB} and \textit{UB} of [1.853, 1.9201] on the \textit{CR}.
\item Classification, characterization of input job sequences and ranking of online scheduling algorithms based on real world inputs.
\item The design of deterministic algorithms for online scheduling on $m$-identical machines have been witnessed various strategies such as \textit{greedy}, \textit{input characterization}, \textit{game tree}, \textit{layering} and job \textit{master}. However, it will be interesting to develop a unified deterministic strategy for scheduling an arbitrary sequence of large jobs.
\item Finding the exact competitiveness achievable by deterministic online algorithms. 
\end{itemize}
\subsection{Non-preemptive, Uniform Related Machines}
\label{subsec: Non-preemptive, Uniform Related Machines}
Offline scheduling in non-identical machines was introduced in late seventies by Hrowitz and Sahni [55]. However, the study of online scheduling in uniform related machines was initiated in year 1993 by Aspnes et al. [15]. They proposed the  deterministic algorithm \textit{ASSIGN-2} and achieved a \textit{UB} of $8$. Algorithm \textit{ASSIGN-2} works on the idea of assigning each incoming job $J_i$ to the lowest speed machine(say $M_1$) as long as $l_1+p_i\leq C_{OPT}$(assuming $l_1$ as the load of $M_1$ before the assignment of $J_i$). As the value of $C_{OPT}$ is not known, algorithm \textit{ASSIGN-2} follows a \textit{doubling strategy}. Initially, a smaller value is chosen for $C_{OPT}$ and later the value of $C_{OPT}$ will be set to twice of its previous value, when the scheduling of job $J_i$ on $M_1$ makes $l_1+p_i > C_{OPT}$. Berman et al. [56] obtained \textit{UB} $5.82$ for large $m$. Further, they proved \textit{LB}s $2.28$ and $2.43$ for $m=6$ and $m=9$ respectively. They proposed an algorithm, which works in phases. In each phase, a sequence of jobs are scheduled. The makespan obtained in each phase is represented through one of the nodes of a graph. They verified the achieved competitive ratios through a computer based search in the graph.\\
Ebenlendr and Sgall [57] proved \textit{LB} $2.56$ for the setting, where the speeds of the machines and the processing times of the jobs are in a geometric sequence. They proposed a new lower bound inequality, which is based on the total number of jobs scheduled and the number of jobs assigned per machine. The \textit{LB} was derived as follows: the behavior of the algorithm is first examined in any of the machines to obtain the \textit{UB} on the number of jobs that can be scheduled on that machine. Basically, this \textit{UB} is a ratio of number of jobs and machine speed. They considered such \textit{UB}s on every machines, which is at least $1$ as in each step of the algorithm only one job can be scheduled. Finally, they obtained the \textit{LB} by assuming the common ratio of the geometric sequence to be 1.\\
Jez et al. [58] obtained \textit{LB}s $2.14$ and $2.31$ for $m=4,5$ respectively. They considered the case, where the processing times of the jobs are in a geometric sequence. They worked on the following idea: choose the speeds of the machines so that two fastest machines can only be utilized by any online algorithm for obtaining the \textit{LB}s by analyzing the possible orders of scheduling the jobs on these machines. They achieved better \textit{LB}s of $2.34$ and $2.46$ for $m=6,9$ respectively. Here, they considered five fastest machines and applied a computer based search strategy to eliminate few of the possible patterns of scheduling the jobs. We now present the summary of all important results on the competitive ratios for non-preemptive scheduling on uniform related machines in Table \ref{tab:Summary of Deterministic  Algorithm for Uniform Related Machines} followed by some non-trivial research challenges.
\begin{table}[h]
\centering
\caption{Summary of Important Competitiveness Results}
\begin{tabular} {ccp{4.7cm}c}
\hline
\textbf{Year} & \textbf{Author(s)} & \textbf{Result(s)} & Bound \\
\hline 
1993 & Aspnes et al. [15] &  $8$ & UB \\
2000 &  Berman et al. [56] &  $5.82$ & UB \\
2000 &  Berman et al. [56] &  $2.28$, $m=6$, $2.43$, $m=9$ & LB \\
2012 &  Ebenlendr and Sgall [57] &  $2.56$ & LB  \\
2013 &  Jez et al. [58] &  $2.14$, $m=4$, $2.31$, $m=5$, $2.462775$, $m=9$ & LB  \\
\hline
\end{tabular}
\label{tab:Summary of Deterministic  Algorithm for Uniform Related Machines}
\end{table}\\
\textbf{Research Challenges:}
\begin{itemize}
\item Minimizing or diminishing the gap between current \textit{LB} and \textit{UB} of [2.56, 5.82] on the \textit{CR}.
\item Development of an alternative to doubling strategy for the improvement of the existing competitive bounds. 
\item Design of efficient competitive online deterministic algorithms with a new parameter or function based on the size of the jobs.  
\item To obtain a \textit{tight} bound for large $m$.
\end{itemize}
\subsection{Preemptive, Identical, Uniform Related Machines}
\label{subsec: Preemptive, Identical and Uniform Related Machines}
Here, we survey important results obtained by deterministic algorithms for preemptive online scheduling in identical and uniform related machines as follows.\\ 
\textbf{Identical Machines.} Less attention has been paid to the online preemptive scheduling on identical machines. To the best of our knowledge, the only deterministic online algorithm \textit{CVW} was proposed by Chen et al. [35] for makespan minimization in identical machines. They obtained \textit{LB} $\frac{m^m}{m^m-(m-1)^m}$ for $m \geq 2$, which tends to $1.58$ for $m \rightarrow \infty$. The overall idea of the algorithm is to maintain the load of the least loaded machine as small as possible so that whenever a large size job arrive it can be assigned to the least loaded machine. The objective is to obtain a bound, which is asymptotically lesser than $2$.\\
\textbf{Uniform Related Machines.} Wen and Du [59] achieved  \textit{UB} $1+\frac{{S_1}.{S_2}}{{S_1}^2+{S_1}.{S_2}+{S_2}^2}$ for $m=2$, where $S_1$ and $S_2$ are the speeds of machine $M_1$ and $M_2$ respectively. Epstein [60] studied a special case, where the speeds of the machines are  defined by the following inequality: $\frac{S_{j-1}}{S_j} \leq \frac{S_J}{S_{j+1}}$, for $j=2......m-1$. He obtained an \textit{UB} for each sequence of speeds as ${\sum_{j=1}^{m}{\frac{S_i}{X}({1-\frac{S_1}{X})^{i-1}}}}^{-1}$, where $X=\sum_{j=1}^{m}{S_i}$.\\ Ebenlendr and Sgall [61] obtained \textit{UB} $4$ by using the doubling strategy for guessing the value of $C_{OPT}$. They showed that algorithm \textit{LS} of Graham [5] achieves \textit{UB} $9+6\log_2{m}$ and \textit{LB} $\log_2{m}$ for preemptive online scheduling on $m$-uniform related machines. Recently, Ebenlendr et al. [62] obtained a \textit{UB}, which lies between the values $2.054$ and $2.718$. They proved that the result is optimal even among all randomized algorithms and any fixed combinations of speed of the machines. We now present some non-trivial research challenges as follows. \\
\textbf{Research Challenges:}
\begin{itemize}
\item Determination of competitive bounds i.e. \textit{UB} for identical machines case and \textit{LB} for uniform related machines case.
\item Design of strategies that avoid or minimize idle periods of uniform related machines while scheduling the jobs online.
\item How to guess the value of the optimum makespan($C_{OPT}$)? In general, as the choices for $C_{OPT}$ go up, the number of cases in the analysis of any online algorithm grow exponentially.
\item Design of nearly optimum online algorithms with best competitive ratios.
\end{itemize}
\section{Randomized Online Scheduling: State of the Art}
\label{sec: Randomized Online Scheduling: State of The Art}
This section is devoted to an overview of the state of the art contributions in design of randomized algorithms for online scheduling with makespan minimization. Important results achieved for non-preemptive online scheduling in identical and uniform related machines are discussed followed by an overview of the seminal works in preemptive online scheduling.
\subsection{Non-preemptive, Identical Machines}
\label{subsubsec: Non-preemptive, Identical Machines}
Design of randomized algorithms for online scheduling  has received notable research interests after the seminal work of Bartal et al. [36]. They proposed the algorithm \textit{Rand-2} for non-preemptive online scheduling on $2$-identical machines. The objective of algorithm \textit{RAND-2} is to maintain an expected load difference of $(\frac{1}{3})L$ between two machines at any instance of time, where $L$ is the sum of processing time of the jobs that have already received. Algorithm \textit{Rand-2} achieves \textit{LB}s of $1.33$ and  $1.4$ for $m=2,3$ respectively.\\
Chen et al. [63] proved that any randomized algorithm for online scheduling on $m$-identical machines with an objective to minimize the makespan must be at least $(\frac{m^m}{m^m-({m-1})^m})$-competitive for all $m$. The bound tends to $\frac{e}{e-1} \approx 1.58$ as $m \rightarrow \infty$. The overall idea of the algorithm is presented as follows: upon receiving a new job $J_i$, the algorithm first computes $m$ probabilities $0\leq x_1,..........x_m \leq 1$, where $\sum_{j=1}^{m}{x_j}=1$. Then, schedule job $J_i$ on machine $M_j$ with probability $x_j$, $j=1,.....m$.\\
Sgall [64] proved that any randomized algorithm \textit{A} must be at least $(1+\frac{1}{(\frac{m}{m-1})^m-1})$-competitive. He showed that the \textit{LB} tends to $1.5819$ as $m \rightarrow\infty$. He defined an interesting ordering of the machines by considering last $m$ jobs of any job sequence. The idea is to order the machines in such a way that the index of the $i^{th}$ machine remains unchanged after the assignment of each new job $J_i$ to it. Initially, when no job is scheduled on any of the machines, then a new job $J_i$ is scheduled arbitrarily on any one of the $m$-machines. The objective of such ordering is to maintain the ratio of the loads of $m$ machines as $1:L:L^2:...\hspace{0.2cm}L^{m-1}$, where $L=\frac{m}{m-1}$. This implies $C_A\geq L^{m-1}$ and $C_{OPT}$=$\frac{1+L+L^2+....+L^{m-1}}{m}=\frac{L^m-1}{m(L-1)}$. Therefore, $\frac{C_A}{C_{OPT}}\geq \frac{L^{m-1}\cdot (m\cdot (L-1))}{L^m-1}\geq 1+\frac{1}{(\frac{m}{m-1})^m-1}$. \\
Seiden [65] proposed the algorithm \textit{LI} by generalizing the 2-machine algorithm of Bartal et al. [36]. Algorithm \textit{LI} schedules an incoming job $J_i$ either on the least loaded machine or the second least loaded machine with certain probability. The objective is to keep the load of one of the $m$ machines as low as possible to schedule on it the largest job that likely to arrive in future. Algorithm \textit{LI} achieves competitive ratios of $1.55665$, $1.65888$, $1.73376$, $1.78295$ and $1.81681$ for $m=3$, $4$, $5$, $6$, $7$ respectively. \\
Albers [44] envisioned a strategy for designing of randomized algorithms by combining simple deterministic policies. She developed the algorithm \textit{RAND}, which is a combination of two deterministic algorithms ${ALG}_1$ and ${ALG}_2$. She obtained an upper bound of $1.916$ for all $m$. Upon the arrival of a job, algorithm \textit{RAND} invokes ${ALG}_i$, $i \in \{1,2\}$ with probability $\frac{1}{2}$, then schedules the whole job sequence by the chosen algorithm. Algorithm \textit{RAND} aims at maintaining two schedules at any instance of time unlike algorithm \textit{RAND-2} and algorithm \textit{LI}, those maintain separate schedules upon each job arrival. Here, Albers proved that none of the known deterministic online strategies can beat the performance of algorithm \textit{RAND}.\\
Tichy [66] obtained \textit{LB}s of 1.425, 1.495 and 1.504 for machines $m=3,5,6$ respectively. Here, a set of algorithms are considered which schedule an incoming job either on the machine with minimum load or on the $k^{th}$ least loaded machine, where $k$ is a positive integer. In [67], he showed the problem for three identical machines case. He considered a critical job sequence which is characterized by some integer parameters and proved by contradiction that no randomized algorithm can obtain a lower bound less than or equal to $1.421$. We now present the summary of important results on the competitive ratios obtained by randomized algorithms for identical machines in Table \ref{tab:Summary of Randomized  Algorithm for Identical Machines}.\\\\
\begin{table}[!ht]
\centering
\caption{Summary of Important Competitiveness Results}
\begin{tabular} {ccp{6.3cm}c}
\hline
\textbf{Year} & \textbf{Author(s)} & \textbf{Result(s)} & \textbf{Bound}  \\
\hline 
1995 & Bartal et al. [36] &  $1.33$, $m$=2 and $1.4$, $m$=3 & LB \\
1994 &  Chen et al. [63] &  $1.58$, $m\rightarrow \infty$ & LB \\
1997 &  Sgall  [64] &  $1.582$ & LB \\
2000 &  Seiden [65] &  $1.55665$, $m$=3,
$1.65888$, $m$=4
$1.73376$, $m$=5, 
$1.78295$, $m$=6, 
$1.81681$, $m$=7 & UB \\
2002 &  Albers [44] &  $1.916$ & UB \\
2002 & Tichy [66] & $1.425$,  $m=3$, $1.495$, $m=5$, $1.504$, $m=6$ & LB \\
2004 & Tichy [67] & $1.421$, $m=3$ & LB \\
\hline 
\end{tabular}
\label{tab:Summary of Randomized  Algorithm for Identical Machines}
\end{table}
\textbf{Research Challenges:}
\begin{itemize}
\item Tighten the gap between the current best lower and upper bounds of $1.582$ and $1.916$ respectively on the competitive ratio for general case of $m$ machines.
\item Development of more refined strategies with stronger invariants to construct improved bounds for the problem.
\item Design of a unified strategy to schedule a sequence of very large \textit{size} jobs. 
\item Development of a new framework to measure the performance of any online scheduling algorithm based on the input characterization approach of Seiden [63] for any job sequence and a general case of $m$ machines.
\end{itemize}
\subsection{Non-preemptive, Uniform Related Machines}
\label{subsec: Non-preemptive, Uniform Related Machines}
Indyk [68] initiated the study of randomized algorithms for non-preemptive online scheduling on uniform related machines. He obtained an upper bound of $5.436$ by extending the work of Aspnes et al. [15].\\
Epstein et al. [43] proposed the algorithm \textit{BIAS} for online scheduling in two uniform related machines with speeds 1 and $S\geq1$ respectively. They obtained a lower bound of $1.5625$ for the speed interval $1 < S < 2$. They showed that randomization does not improve the bounds for the speeds $S\geq2$.\\
Berman et al. [69] proposed a $4.311$-competitive algorithm by refining the doubling strategy of Aspnes et al. [15]. Also, they achieved a lower bound of $1.837$. The algorithm works in the following way: an initial guess for a smaller value of the overall optimum load $C_{OPT}$ is made along with the initialization of variables $c_j$, $x_j$ and $r$. Here, $c_j$ is referred to as the capacity of machine $M_j$ which is the amount of work machine $M_j$ can do under load $C_{OPT}$. We have $c_j=C_{OPT}(S_j)$ for any $M_j$, where $S_j$ is the speed of machine $M_j$. The variable $r$ is initialized to $\sqrt(\frac{1}{2})$. Upon receiving a job, first the value of $C_{OPT}$ is checked. The value of $C_{OPT}$ is updated if \textit{OnlyFor}($S_j$, $C_{OPT}$, $\sigma)> C_{OPT}.Cap(S)$ for some $S \in \{S_1, S_2,....S_m\}$.(Here, $\sigma$ is referred to as a sequence of jobs and OnlyFor($S_j$, $C_{OPT}$, $\sigma$) defines the sum of sizes of those jobs for which $\frac{p_i}{S_j} > C_{OPT}$ for some $j \in \{j=1, 2,...m\}$ ). Then, the value of $C_{OPT}$ is updated to $r(C_{OPT})$. The values of $x_j$ and $c_j$ are updated as follows: $x_{j+1}=(C_{OPT})S_j$ and $c_{j+1}=x_{j+1}+C_{OPT}$. Here, the value of $r$ is chosen uniformly at random from the interval $[-z, 1-z]$. For a negative $z$, we have $r=r^{y+1}$ and for a positive $z$, we have $r=r^y$ with any integer variable $y$. The main idea of the algorithm is to increase the value of $C_{OPT}$ at each phase by a factor $r$ instead of doubling its value as soon as the current value of $C_{OPT}$ seems to be very small. Hence, the jobs arriving upfront can be scheduled on the relatively faster machines and the jobs coming later in the sequence can be judiciously assigned to the slower machines. Therefore, the loads of the faster machines can be increased, which in turn minimizes the overall makespan of the schedule.            \\
Epstein and Sgall [70] obtained an improved lower bound of $2$ by characterizing the input parameters such as number of jobs, processing times of the jobs, number of machines and their speeds. For an infinite  number of machines, where a machine $M_j$ has speed $S_j=x^j$ for any variable $x$<1, they obtained a lower bound of $1+x$. Here, they considered an infinite job sequence, where the processing time $p_i=x^i$ for any job $J_i$ such that the sum of processing times of all jobs is always $\frac{x}{1-x}$. For a general case of $m$ machines, they obtained a simple lower bound of $\frac{1+x}{1+x^m}$ by considering $m$ largest jobs each with processing time $p_i=x^i$ and machines with speed $S_j=x^j$, where $0<x<1$ such that total processing time of the jobs is always $\frac{x(1-x^m)}{1-x}$. Then, by judiciously modifying the values of input parameters, they obtained improved lower bounds of $1.33$ and $1.96234$ for $m=2, 100$ number of machines respectively. We now present the summary of important competitive results in non-preemptive online scheduling on uniform related machines by randomized algorithms in Table \ref{tab:Summary of Important Results for Uniform Related Machines by Randomized Strategies}.
\begin{table}[h]
\centering
\caption{Summary of Important Competitiveness Results}
\begin{tabular} {ccp{6.2cm}c}
\hline
\textbf{Year} & \textbf{Author(s)} & \textbf{Result(s)} & \textbf{Bound} \\
\hline 
1997 & Indyk [68] & $5.436$ & UB  \\
2000 &  Berman et al. [69] &  $4.311$ & UB   \\
2000 &  Berman et al. [69] &  $1.8372$ & LB \\
2000 &  Epstein, Sgall [70] &  $2$ & LB\\

2001 & Epstein et al. [43] & $1.5625$ & LB \\
\hline
\end{tabular}
\label{tab:Summary of Important Results for Uniform Related Machines by Randomized Strategies}
\end{table}\\
\textbf{Research Challenges:}
\begin{itemize}
\item Close or diminish the gap of [2, 4.311] for the current best lower and upper bounds on the competitive ratio.
\item Design of competitive randomized online algorithms with a new function and parameter based on the speeds of the machines.
\item Can a randomized algorithm beat the performance of the current best deterministic strategy in obtaining the lower bound?
\end {itemize}  
\subsection{Preemptive, Identical, Uniform Related Machines}
\label{subsec: Preemptive, Identical, Uniform Related Machines}
Here, we present an overview of the state of the art literature in design of randomized algorithms for preemptive online scheduling in identical and uniform related machines as follows.\\
\textbf{Identical Machines.} Seiden [65] proposed the first randomized algorithm for preemptive online scheduling on identical machines by slightly modifying the original algorithm of Chen et al.[63]. Seiden considered the notion of job splitting as the preemptive characteristic of the scheduling algorithm. The algorithm schedules a sequence of jobs on $m$ identical machines in the following way: whenever a job $J_i$ is received, a time slot is assigned for job $J_i$ on each machine starting from the most loaded machine $M_m$ to the least loaded machine $M_1$. The time slot for $J_i$ in machine $M_m$ is defined as a function($f_1$) of the following two parameters: sum of the processing times of all jobs(T) and largest processing time($P_b$). Formally, the time slot in $M_m$ can be $(L_m, f_1(T,P_b)]$. The time slot for job $J_i$ on rest of the machines($j<m$) are allocated in the following way: $(L_j, L_{j+1}]$. The objective is to keep $k$ machines lightly loaded and $m-k$ machines heavily loaded. Here, the bounds on the competitive ratios were shown as a function of real constants and were verified through computer programs written by tool \textit{Mathematica}. However, a general bound for the algorithm was not derived.\\
\textbf{Uniform Related Machines.} Ebenlendr and Sgall [61] obtained an upper bound of $2.71$ for preemptive online scheduling on uniform related machines by using the doubling strategy. They made a conjecture on the improvement over doubling strategy that the initial value for optimal makespan($C_{OPT}$) can be guessed by considering an exponential distribution.\\
Epstein and Sgall [70] obtained lower bounds on the competitive ratio for the worst case combination of speeds for any fixed $m$. These bounds approach to $2$ when $m\rightarrow \infty$ and all hold for randomized algorithms. 
\section{Recent Trends}\label{sec:Recent Trends}
Online scheduling poses a non-trivial research challenge for designing optimal algorithms due to unavailability of complete input information at the outset. Recent studies have guaranteed a performance improvement over the online deterministic and randomized strategies by relaxing one or more stringent constraints of the pure online scheduling setting. The relaxation includes availability of additional computational power or \textit{extra piece of information(EPI)} to an online scheduling algorithm. This section highlights some of the relaxed variants of the online scheduling model and reports important competitive analysis results as follows.    
\subsection{Resource Augmentation}\label{subsec:Resource Augmentation}
Resource augmentation model was pioneered by Kalyansundaram and Pruhs [71]. Here, an online scheduling algorithm is given some additional resources such as \textit{high speed machines(speed augmentation)} or \textit{memory space (memory augmentation)} as compared to its \textit{optimum offline(OPT)} counterpart. We now discuss on speed augmentation and memory augmentation as follows.
\subsubsection{Speed Augmentation}\label{subsubsec:Speed Augmentation} 
This model gives as input a set of relatively high speed machines to an online algorithm for scheduling a sequence of jobs and compare its performance with \textit{OPT} that schedules jobs on relatively slower machines. For instance, an online scheduling algorithm is given a set of machines with speeds $S_j\geq 1$, $\forall M_j$, where algorithm \textit{OPT} operates machines with speed $S_j$=$1$, $\forall M_j$. Berman and Coulston [72] studied preemptive online scheduling on a single machine with speed augmentation for minimizing $\sum_{i=1}^{n}{c_i-r_i}$, where $r_i$ is the release time of job $J_i$ and $c_i$ is the completion time of $J_i$. They considered that an online algorithm is given a machine with speed $u$ times faster than the machine given to the optimal offline algorithm. They proposed the algorithm \textit{Balance}, which always schedules least executed job. They achieved a \textit{UB} of $\frac{u}{u-1}$. Algorithm \textit{Balance} was shown to be $(\frac{2}{u})$-competitive for $u\geq 2$. Lam and To [73] studied preemptive online scheduling with hard deadlines. They explored a trade-off between increment of speed and number of machines. They proved that any online algorithm that schedules jobs by earliest deadline(EDF) is optimal for $m$=$2$, if the algorithm is given machines with speeds $1.5$ times faster than its optimal offline counterpart. It was shown that \textit{EDF} rule achieves optimality with $(2-\frac{1+x}{m+x})$ times faster machines and $x\geq 0$ additional machines. Some recent works on online scheduling with speed augmentation can be found in [74-76].
\subsubsection{Memory Augmentation}\label{subsubsec:Memory Augmentation}
\textbf{Buffer}. A buffer $B(k)$ is given, which is capable of keeping at most $k$ jobs, where $k\geq 1$. Availability of $B(k)$ allows an online algorithm either to keep an incoming job temporarily on the buffer or to schedule one of the available jobs directly on a machine. An online algorithm now can see at most $k+1$ jobs at any time step prior to make a scheduling decision. Some of the interesting studies for online scheduling in identical machines with buffer of varying sizes are due to [77-79, 81]. In [82], the authors considered online \textit{hierarchical} scheduling with $B(1)$ in $2$-identical machines, where the machines are of different capabilities in the sense that machine $M_1$ can process any job and machine $M_2$ can process only some designated jobs. An available job $J_i$ is given with its $p_i$ and $g_i$, if $g_i$=$1$, then $J_i$ can only be processed by machine $M_1$, if $g_i$=$2$, then $J_i$ can be processed by either of the machines. We now represent important competitive analysis results achieved for online scheduling in identical machines with buffer in Table \ref{tab:Important Results for Online Scheduling with Buffer}.
\begin{table}[!ht]
\centering
\caption{Important Results for Online Scheduling with Buffer}
\begin{tabular} {ccc}
\hline
\textbf{Year, Author(s)} & \textbf{Machine(s), $B(k)$} & \textbf{Competitive Ratio(s)} \\
\hline 
1997, Kellerer et al. [77] & $2$-identical, $k\geq 1$ &  $1.33$ Tight \\
1997, Zhang [78] & $2$-identical, $k$=$1$  &  $1.33$ Tight\\
2004, Dosa, He [79] &  $2$-identical, $k$=$1$ &  $1.25$ Tight\\
2012, Lan et al. [81] &  $m$-identical, $k$=$(1.5)m$ &  $1.5$ Tight\\
2013, Chen et al. [82] & $2$-identical, hierarchical, $k$=$1$ &  $1.5$ Tight  \\
\hline 
\end{tabular}
\label{tab:Important Results for Online Scheduling with Buffer}
\end{table}\\
\textbf{Parallel Schedules}. Upon receiving a job $J_i$, $1\leq i\leq n$, an online algorithm makes two copies of $J_i$ and virtually schedules each of the copies by two independent procedures. Therefore, two parallel schedules are constructed at any time step. After constructing two virtual schedules for the entire job sequence, one of the schedules is chosen that has incurred minimum $C_{max}$ for actual assignment of all jobs. Here, an extra space is given to maintain solutions for parallel schedules. In 1997, Keller et al. [77] first studied non-preemptive online scheduling on $2$-identical machine with parallel schedules and obtained a \textit{tight} bound of $1.33$. In 2012, Albers and Hellwig [83] investigated the general case of $m$-identical machine and achieved \textit{UB} $1.75$. An open issue is to obtain a \textit{tight} bound for $m$-identical machine setting. It will be interesting further to explore uniform machine setting with parallel schedules.
\subsection{Semi-online Scheduling}\label{subsec:Semi-online Scheduling}
The study of online scheduling with \textit{Extra Piece of Information(EPI)} pioneered the concept of semi-online scheduling. Kellerer et al. [77] first envisioned that availability of additional information on future inputs is quite natural in contrast to the constraint of no information at all. For an instance, the number of jobs that are going to be submitted to a multi-user time shared system is not known a priori. However, the minimum and maximum time required to process each job can be known in advance by previous history. This revitalized the area of online scheduling to explore practically significant new \textit{EPI}s. We now report important results, achieved by some well-known semi-online policies with a classification of \textit{EPI}s as follows.
\subsubsection{Sum}\label{subsubsec:Sum} 
An online algorithm is given a priori, the sum of the processing times of all jobs. Kellerer et al. [77] first introduced \textit{Sum} as an \textit{EPI} for non-preemptive online scheduling on $2$-identical machine and achieved a \textit{tight} bound of $1.33$ on the \textit{CR}. Recent contributions in this setting are due to [80, 84, 86-92]. Important results achieved for online scheduling with known \textit{Sum} is reported in Table \ref{tab:Important Results for Online Scheduling with Known Sum}.
\begin{table}[!ht]
\centering
\caption{Important Results for Online Scheduling with Known Sum}
\begin{tabular} {ccp{5.8cm}}
\hline
\textbf{Year, Author(s)} & \textbf{Machine(s)} & \textbf{Competitive Ratio(s)} \\
\hline 
1997, Kellerer et al. [77] & $2$-identical &  $1.33$ Tight \\
1998, Gilrich et al. [84] & $m$-identical &  $1.66$ UB\\
2004, Angelleli et al. [86] &  $m$-identical &  $(1.565, 1.725)$ LB and UB respectively, $m\rightarrow \infty$\\
2005, Cheng et al. [87] &  $m$-identical &  $(1.5, 1.6)$ LB and UB respectively, $m\geq 6$\\
2007, Angelleli et al. [88] & $3$-identical &   $(1.392, 1.421)$ \textit{LB} and \textit{UB} respectively \\
2008, Angelleli et al. [80] & $2$-uniform & $1.33$ Tight for $s$=$1$, $(1+\frac{1}{s+1})$ Tight for $s\geq 1.732$  \\
2009, Ng et al. [89] & $2$-uniform & $1.369$ UB \\
2010, Angelleli et al. [90] & $2$-uniform & $1.359$ LB for $s$=$1.5$\\
2011, Dosa et al. [91] & $2$-uniform & $1+\frac{1}{3s}$ Tight\\
2015, Keller et al. [92] & $m$-identical & $1.585$ Tight\\
\hline 
\end{tabular}
\label{tab:Important Results for Online Scheduling with Known Sum}
\end{table}
\subsubsection{Optimum Makespan}\label{subsubsec:Optimum Makespan} 
An online algorithm is given with the value of the \textit{optimum makespan(Opt)} for an online sequence of jobs prior to their scheduling and availability. Epstein [93] first considered \textit{Opt} as an \textit{EPI} for online scheduling on $2$-uniform machine and obtained \textit{UB} $1.414$. Earlier, Azar and Regev [94] introduced \textit{Opt} as an \textit{EPI} in bin stretching problem and achieved \textit{UB} $1.625$. The latest results have been contributed in [89, 91, 95, 96]. Important results achieved for online scheduling with known \textit{Opt}is reported in Table \ref{tab:Important Results for Online Scheduling with Known Opt}.
\begin{table}[!ht]
\centering
\caption{Important Results for Online Scheduling with Known Opt}
\begin{tabular} {ccp{6.1cm}}
\hline
\textbf{Year, Author(s)} & \textbf{Machine(s)} & \textbf{Competitive Ratio(s)} \\
\hline 
1998, Azar, Regev [94] & $m$-identical &  $1.625$ UB \\
2003, Epstein [93] & $2$-uniform &  $1.414$ UB\\
2009, Ng et al. [89] &  $2$-uniform &  $1.366$ Tight \\
2011, Dosa et al. [91] &  $2$-uniform &  $\min\{1+\frac{1}{3s}, 1+\frac{3s}{5s+5}, 1+\frac{1}{2s+1}\}$ LB for $s\geq 1$ \\
2015, Dosa et al. [95] & $2$-uniform &   $\frac{6(s+1)}{4s+5}$ LB for $s\in [1.395, 1.443]$, $\min\{\frac{12s+10}{9s+7}, \frac{18s+16}{16s+7}, \frac{8s+7}{3s+10}\}$ LB for $s\in [1.66, 1.72]$ \\
2017, Dosa et al. [96] & $2$-uniform & $\frac{2s+10}{9s+7}$ Tight for $s$=$1.725$, $\frac{s+1}{2}$ Tight for $1.725\leq s\leq 1.732$\\
\hline 
\end{tabular}
\label{tab:Important Results for Online Scheduling with Known Opt}
\end{table}
\subsubsection{Max}\label{subsubsec:Max} 
A job with largest processing time(\textit{Max}) is known at the outset. He and Zhang [97] introduced \textit{Max} as known \textit{EPI} in online scheduling on $2$-identical machines and obtained a \textit{LB} of $1.33$. Further studies for online scheduling with known \textit{Max} are due to [98-101]. Important results achieved for online scheduling with known Max is reported in Table \ref{tab:Important Results for Online Scheduling with Known Max}.
\begin{table}[!htbp]
\centering
\caption{Important Results for Online Scheduling with Known Max}
\begin{tabular} {ccp{5.8cm}}
\hline
\textbf{Year, Author(s)} & \textbf{Machine(s)} & \textbf{Competitive Ratio(s)} \\
\hline 
1999, He, Zhang [97] & $2$-identical &  $1.33$ Tight \\
2002, Cai [98] & $m$-identical &  $1.414$ UB\\
2004, He, Jiang [99] &  $2$-uniform &  $\frac{2s^2+3s+1}{2s^2+2s+1}$ LB for preemptive scheduling\\
2008, Wu et al. [100] &  $m$-identical &  $2-\frac{1}{m-1}$ Tight \\
2013, Lee, Lim [101] & $m$-identical &   $1.618$ Tight for $m$=$4$ and $1.667$ Tight for $m$=$5$  \\
\hline 
\end{tabular}
\label{tab:Important Results for Online Scheduling with Known Max}
\end{table}
\subsubsection{Tightly Grouped Processing Time(TGRP)}\label{subsubsec:Tightly Grouped Processing Time(TGRP)}
Processing times of an online sequence of jobs are not known a priori. However, maximum and minimum time required to process each job is given to an online algorithm at the outset. For instance, it is given that size of each job $J_i$, ($1\leq i\leq n$) is in between $(p, rp)$, where $p>0$ and $r\geq 1$. He and Zhang [97] initiated study of online scheduling on $2$-identical machine with $TGRP(p, rp)$ and achieved \textit{LB} $1.33$. Further advancements in this setting are the outcomes of the following contributions [99, 102]. Later, we shall report some of the results, where \textit{minimum processing time$(TGRP(lb))$} for each job or \textit{maximum processing time$(TGRP(ub))$} for each job were considered in combine with some other \textit{EPI}s. Recently [103], online hierarchical scheduling on $2$-uniform machine with $TGRP(1, \alpha)$ has been studied and a \textit{LB} of $1+\alpha$ has been shown, where $\alpha\geq 1$.  Important results achieved in the literature for online scheduling with \textit{TGRP} is reported in Table \ref{tab:Important Results for Online Scheduling with Known TGRP}.
\begin{table}[!ht]
\centering
\caption{Important Results for Online Scheduling with Known TGRP}
\begin{tabular} {ccp{5.2cm}}
\hline
\textbf{Year, Author(s)} & \textbf{Machine(s)} & \textbf{Competitive Ratio(s)} \\
\hline 
1999, He, Zhang [97] & $2$-identical &  $1.5$ LB \\
2004, He, Jiang [99] & $2$-uniform &  $\max\{\frac{1+s+r/2+rs/2}{1+s+rs/2}, \frac{s^2+s}{s^2+1}\}$ Tight for $s>\sqrt{2}$ and $s\leq r< \frac{2s^2-2}{s}$ with preemptive scheduling\\
2005, He, Dosa [102] &  $3$-identical &  $1.5$ UB for $r\in (2, 2.5)$, $\frac{4r+2}{2r+3}$ UB for $r\in (2.5, 3)$\\
2015, Luo, Xu  [103] &  $2$-uniform, hierarchical &  $(1+\alpha)$ LB \\
\hline 
\end{tabular}
\label{tab:Important Results for Online Scheduling with Known TGRP}
\end{table}
\subsubsection{Arrival Sequence of the Jobs}\label{subsubsec:Arrival Sequence of the Jobs}
An online algorithm is given with the arrival pattern of a sequence of jobs. For an instance, it is known at the outset that the jobs arrive one by one with \textit{non-decreasing sizes (Decr)}. Seiden et al. [104] introduced \textit{Decr} as a known \textit{EPI} for preemptive online scheduling on $2$-identical machine and achieved a \textit{tight} bound of $1.16$. For $m$=$3$, \textit{LB} $1.18$ was shown and for a general case of $m$-identical machine, \textit{LB} $1.36$ was obtained. Recent works for online scheduling with known \textit{Decr} can be found in [105, 106] and we report important results on competitive analysis in Table \ref{tab:Important Results for Online Scheduling with Known Decr}.
\begin{table}[!htbp]
\centering
\caption{Important Results for Online Scheduling with Known Decr}
\begin{tabular} {ccp{5.1cm}}
\hline
\textbf{Year, Author(s)} & \textbf{Machine(s)} & \textbf{Competitive Ratio(s)} \\
\hline 
2000, Seiden et al. [104] & $m$-identical &  $1.36$ LB for $m\rightarrow \infty$, $1.16$ Tight for $m$=$2$, $1.18$ LB for $m$=$3$ \\
2005, Epstein, Favoholdt [105] & $2$-uniform &  $1.28$ Tight\\
2012, Cheng et al. [106] &  $m$-identical &  $(1.18, 1.25)$ Tight for $m$=$3$ and $m>3$ respectively\\
\hline 
\end{tabular}
\label{tab:Important Results for Online Scheduling with Known Decr}
\end{table} 
\subsubsection{Combined EPIs}\label{subsubsec:Combined EPIs}
An online algorithm is given with more than one \textit{EPI}s at the outset. Angelelli [85] initiated the study of online scheduling on $2$-identical machine with combined \textit{EPI}s. They considered an online algorithm has the prior knowledge of minimum  processing time for all jobs and sum of the sizes of the jobs. He achieved a \textit{tight} bound of $1.33$ on the \textit{CR}.  Tan and He [107] considered two combined \textit{EPI}s. First, they considered \textit{Sum}, \textit{Max}  and obtained a \textit{LB} of $1.2$. Secondly, they obtained a \textit{LB} of $1.11$ with known \textit{Sum} and \textit{Decr}. Recent contributions in this setting are due to [108-113]. Important results achieved in the literature for online scheduling with combined \textit{EPI}s are reported in Table \ref{tab:Important Results for Online Scheduling with Combined EPIs}.
\begin{table}[!ht]
\centering
\caption{Important Results for Online Scheduling with Combined EPIs}
\begin{tabular} {cp{2cm}p{5.4cm}}
\hline
\textbf{Year, Author(s)} & \textbf{Machine(s), EPIs} & \textbf{Competitive Ratio(s)} \\
\hline 
2000, Angelleli [85] & $2$-identical, Sum, $TGRP(lb)$ &  $1.33$ Tight \\
2002, Tan, He [107] & $2$-identical, Sum, Max & $1.2$ Tight\\
2003, Angelleli et al. [108] &  $2$-identical, Sum, $TGRP(ub)$ &  $1.2$ Tight for $ub\in (0.5, 0.6)$, $1+\frac{ub}{3}$ Tight for $ub\in (0.75, 1)$\\
2006, Angelleli et al. [109] &  $2$-identical, Sum, $TGRP(ub)$ &  $(1+\frac{1}{2b+1})$ Tight for $ub\in [\frac{1}{b}, \frac{2(b+1)}{b(2b+1)}]$, $(\frac{b-1}{3})ub$+$0.66(\frac{b+1}{b})$ Tight for $ub\in (\frac{2b-1}{2b(b-1)}, \frac{1}{b-1}]$ and $b\geq 2$ \\
2006, Hua et al. [110] & $3$-identical, Sum, Max & $(1.33, 1.4)$ LB and UB respectively\\
2007, Wu et al. [111] & $3$-identical, Sum, Max & $1.33$ Tight\\
2012, Cao et al. [112] & $2$-identical, Opt, Max & $1.2$ Tight\\
2016, Cao, Wan [113] & $2$-identical, $TGRP(1, r)$, Decr & $1.16$ Tight for $1\leq r< 1.5$, $1.16$ LB for $r\geq 1.5$\\
\hline 
\end{tabular}
\label{tab:Important Results for Online Scheduling with Combined EPIs}
\end{table}
\subsubsection{Inexact Partial information}\label{subsubsec:Inexact Partial information}
 Here, the \textit{EPI} given to an online algorithm is not exact. For instance, the algorithm knows a nearest value of actual \textit{Sum} but not the exact value. Tan and He [114] initiated the study of online scheduling in $m$-identical machines with inexact \textit{EPI}s. They considered independently inexact $C_{OPT}$, Sum, Max and obtained lower bound of $1.5$ for each case.\\  
A detail survey on semi-online scheduling for makespan minimization  is presented in [30].
\section{Open Problems(OP)}
\label{sec: Open Problems}
\begin{itemize}
\item  ${OP}_1$: \textbf{Defining a new performance measure.} \\
Can we come up with an alternate performance measure than competitive analysis for online scheduling algorithms? The reason is quite intense in the sense that comparing the cost of an online scheduling algorithm with an actual lowest possible cost would be more realistic than comparing with the lowest cost obtained by an unrealistic offline scheduling algorithm. Now, the non-trivial challenge is to define the actual lowest possible cost.
\item ${OP}_2$: \textbf{Fairness criteria.}\\
Is an online scheduling algorithm \textit{fair} in sharing resources such as machines and time? It is worth of considering \textit{fairness} in online scheduling for multi user systems, where the main concern is to share the resources fairly among the users. Here, a scheduling algorithm mainly focuses on optimizing the objective for each user than the overall objective of the system. Now, the question is: how to analyze the performance of such on line scheduling algorithms?
\item ${OP}_3$: \textbf{Realistic Job Characteristics as new EPIs.}\\
It is an open issue to explore the realistic job characteristics that can be known in advance for the improvement of the existing bounds obtained by online scheduling algorithms in multiprocessor systems.
\item ${OP}_4$: \textbf{Generic unified online scheduling model.}\\
Can we define an unified model for illustrating all variants of the online scheduling? It will be interesting to design a generic online algorithm based on the model that can be applicable for all settings of the online scheduling problem. 
\item ${OP}_5$: \textbf{Characterization of input job sequences for practical applications.}\\
How to characterize input job sequences? Characterization of input sequences will provide a mapping rule for the practitioner to implement various theoretical  online scheduling strategies in real world applications.    
\end{itemize}
\section{Concluding Remarks}
We presented the state of the art results for preemptive and non-preemptive online scheduling with makespan minimization. Important contributions on deterministic and randomized online algorithms in parallel machine models such as identical and uniform-related were discussed. The basic concepts of online algorithm, optimum offline algorithm and competitive analysis were presented from a beginner's perspective.  Well-known related previous surveys on online scheduling for the last five decades were summarized in a chronological way. Fourteen well-known online scheduling algorithms along with their competitive analysis results were presented.\\
Two emerging research trends such as resource augmentation in online scheduling and semi-online scheduling with extra piece of information were also highlighted. We explored non-trivial research challenges and open problems in our survey.  We hope that our survey will help the naive researchers to gain a basic and comprehensive understanding of the emerging area of online scheduling with makespan minimization and inspire for future research.\\\\
\textbf{Acknowledgment}\\\\
This work is partially supported by Department of Computer Science and Engineering of Veer Surendra Sai University of Technology, Burla, Sambalpur, India. \\\\





%
%
%
\textbf{References}\\

\end{document}